\documentclass[12pt]{article}
\usepackage{graphicx}

\begin{document}

\title{On the capabilities of survey telescopes \\ of moderate size}
\author{V.~Yu.~Terebizh$^{1,2}$\thanks{E-mail: valery@terebizh.ru} \\
$^{1}$Crimean Astrophysical Observatory, Nauchny, Crimea 298409 \\
$^{2}$Institute of Astronomy RAN, Moscow 119017}

\date{February 09, 2016}

\maketitle

\begin{abstract}
To explore capabilities of moderate-size optical telescopes in surveys, 
the set of 9 new wide-field designs having apertures up to 1~m is considered. 
All but one systems have angular field of view in a range $3.5^\circ - 10^\circ$ 
and flat focal surface; the field of the last system is $45^\circ$ in diameter at 
the 0.5~m aperture and spherical focal surface. The complete description of the 
optical layouts is given in the Appendix. Relations between the expected limiting 
magnitude, survey speed and exposure time allow to choose the system that is most 
suitable for a particular task of observations. In principle, a single wide-field 
telescope with the aperture of approximately 1~m can detect objects brighter than 
$22.5^m$ over the entire hemisphere within one night, however, the reliability of 
acquired data can be significantly increased by using a hierarchic observational 
network comprised of telescopes with optimized parameters.  
\end{abstract}

\section{Introduction}

A number of important astrophysical problems necessitates continuous registration
of all objects of the sky brighter than about $23^m$ in the visible waveband. 
To estimate the required survey speed $S$, measured in square degrees per second 
(deg$^2$/sec), we imply that one needs to cover a sky area of $10^4$ deg$^2$ within 
3~hours. This area is a little smaller than the entire hemisphere visible above the 
horizon and free of absorption in the Milky Way and Earth's light pollution at large 
zenith distances. This gives us the value of $S \simeq 1$ deg$^2$/sec and makes the
problem quite non-trivial. Indeed, the field of view of a so-called {\em classical 
telescope} (parabolic primary mirror plus hyperbolic secondary mirror) is only 
several arc minutes wide. Thus, one needs to acquire about $10^6$ images to cover 
the considered sky area, which is unrealistic even with several telescopes. 
Within this scope even Ritchey-Chretien telescopes, recently considered wide-field, 
fail to solve the problem: the typical field of a Ritchey-Chretien telescope 
does not exceed $20'$, which might reduce the number of images, mentioned above, 
by only an order of magnitude. 

In this regard, we would naturally turn to a remarkable Bernhardt Schmidt (1930) 
system, whose modified versions reach a field of about $10^\circ$ in diameter. 
Astronomers had used these systems for over fifty years while the photographic 
plates were light detectors: one had to bend the plate to match the curved focal 
surface of the Schmidt camera (its curvature radius is about the effective focal 
length). Meanwhile, the majority of modern detectors are flat. One can achieve 
the flat field either by complicating the optical system or by making the field 
faceted with small field-flatteners. The last option has been applied to the 
{\it Kepler} telescope, which has the aperture of $95$~cm and the field area of 
$115$ deg$^2$. Its detector consists of 21 pairs of ordinary 59~mm $\times$ 28~mm 
CCDs covered by sapphire field-flattening lenses. Obviously, this way is feasible 
now only in unique projects, so the most designs discussed below have flat focal 
surface. The exception is an all-spherical design with a $45^\circ$ field~-- 
the particular implementation of a system, which was proposed recently by the 
author (Terebizh 2015, 2016). 

The first step towards simplification of the Schmidt camera its author made 
himself by testing in 1934 a model with three spherical lenses instead of the 
aspheric corrector (see Wachmann 1955; Busch et al. 2013). In fact, all the 
subsequent wide-field catadioptric telescopes~-- the systems by Richter and 
Slevogt (1941), Schmidt-Houghton (Houghton 1942, 1944), Hawkins and Linfoot (1945), 
Baker (1962) and $\Omega_{2-3}$ (Terebizh 2007a,b)~-- are the successors of the 
two generic systems, invented by Schmidt. Modern versions of these systems provide 
angular field up to $10^\circ$ at flat focal surface and aperture reaching 
1~m (Terebizh 2011).

Another approach, implying a lens corrector mounted in the vicinity of the focus 
of a large aspheric mirror, was introduced by Sampson (1913), Ross (1935) and Wynn 
(1968). Within this approach one cannot achieve a field, comparable with that of 
a mid-size catadioptric system, but the large aperture diameter allows 
detecting faint objects. The modern types of prime-focus correctors were proposed 
by Terebizh (2003) and Saunders et al. (2014). 

This paper was initiated by researchers questions that arise in the development 
of survey projects, first of all: {\em what an optical system is best suited at 
the specified survey depth and speed?} We compare the relevant efficiency of 
various catadioptric designs with the angular field in a range of $3.5^\circ - 
10^\circ$ (flat focal surface) and the $45^\circ$-design with a spherical focal 
surface (Fig.~1). Apertures of catadioptric designs are in the range from 
$0.4$~m up to $1.0$~m. A 20-cm refractive lens with a $15^\circ$ angular field 
and flat focal surface was added for comparison\footnote{In calculations, we 
used the {\it Zemax} optical program (ZEMAX Development Corporation, U.S.A.).}. 

To be specific, the linear diameter $B$ of the flat field has been adopted the 
same for all the designs, $134.5$ mm, which coincides with the length of the 
diagonal CCD STA 1600 of Semiconductor Technology Associates. Thus, we leave 
aside the telescopes with huge mosaic detectors. Obviously, the latter 
provide larger angular field of view, however, our goal now is to maintain the 
uniformity of results. 

Observations with the curved focal surface need special discussion, which we 
give in Section 2.2.

Since one should take into account the specifications of light detectors when 
designing an optical system, we briefly touch on issues related to the matching 
the telescope's and detector's resolving power.

\section{Optical layouts}

\subsection{Flat focal surface}

The set of flat-field systems that we consider hereafter includes a telescope 
with a prime-focus corrector, a corrected Cassegrain system, a Schmidt camera, a 
Schmidt-Houghton telescope, a modified Richter-Slevogt telescope (Terebizh 2001), 
a version of the Amon, Rosin and Jackson camera (Amon et al. 1971), $\Omega_2$ 
and $\Omega_3$ systems (Terebizh 2007a,b), and a lens objective to compare its 
efficiency with that of catadiopric systems (Fig.~1). Basic optical layouts of 
these systems have long been known, almost all of them are implemented. New 
versions were designed for this paper, which provide the widest possible field 
of view given the image quality and size of the detector. 

\begin{figure*}[h]   
   \centering
   \includegraphics[width=140mm]{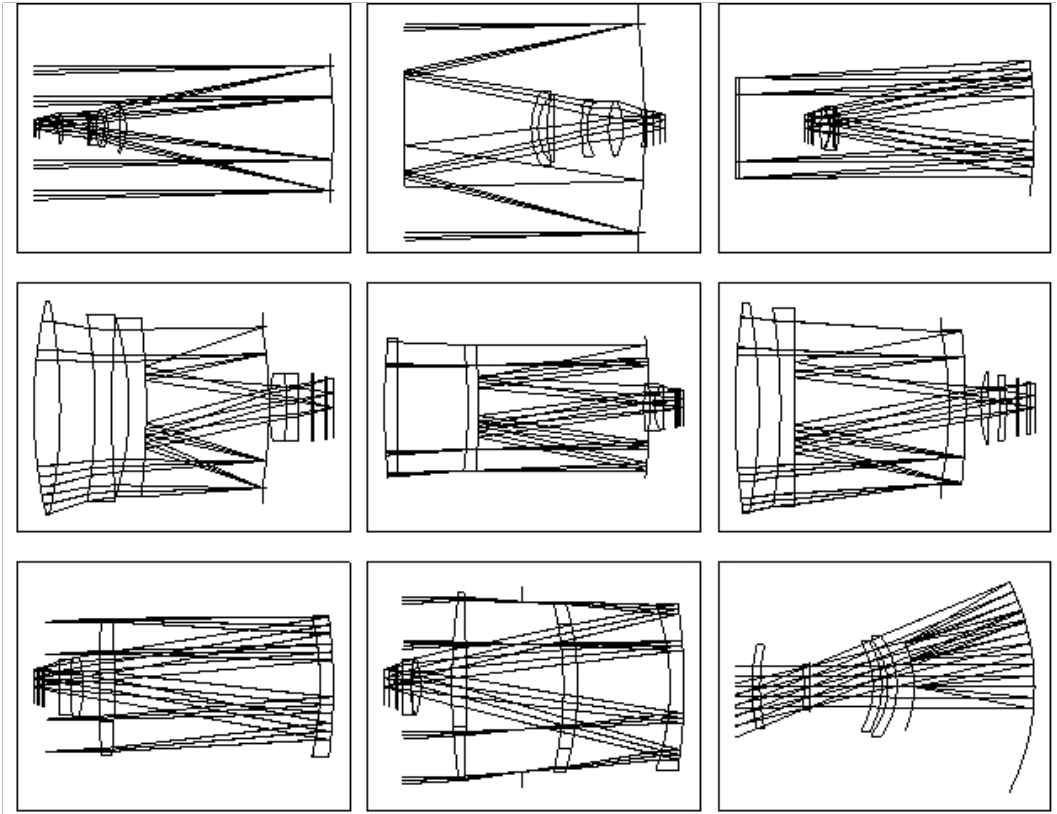}
   \caption{Optical layouts of survey telescopes discussed in this paper. 
   First row, from left to right: VT-56y, -112m, -110f. 
   Second row: VT-78e, -77i, -98v. Third row: VT-102j, -60g, -119j.} 
\end{figure*} 

\begin{table}[h]    
{\footnotesize 
\caption{General characteristics of telescope designs a with flat (No.~1 - 9) and curved (No.~10) focal surface.}
\begin{tabular}{rrrrrcccccr}
\hline \noalign{\smallskip}
 No.&  VT- & D    & $2w^\circ$ & F  & Waveband & U & $D_{80}$ & $D_{80}''$ & $\Gamma$ & L   \\
    &  No. &(mm)  &            &(mm)& ($\mu$m) &   & ($\mu$m) &            & (H)      &(mm) \\
\hline
\noalign{\smallskip}
 1 &  56y & 1000 &  3.5 & 2183 & 0.40-0.85 & 0.85-0.85 &  6.5-7.0  & 0.61-0.66 & 1.90 & 2299 \\
 2 & 112m & 1000 &  3.5 & 2189 & 0.45-0.85 & 0.69-0.69 & 10.6-11.4 & 1.0-1.1   & 1.28 & 1236 \\
 3 & 110f &  500 &  7.5 & 1025 & 0.42-0.82 & 0.75-0.75 &  8.2-10.8 & 1.6-2.2   & 0.71 & 1500 \\
 4 &  78e &  400 & 10.0 &  764 & 0.45-0.85 & 0.72-0.51 &  8.9-10.9 & 2.4-2.9   & 0.47 &  679 \\
 5 &  77i &  500 &  5.0 & 1538 & 0.45-0.85 & 0.74-0.57 & 10.4-13.9 & 1.4-1.8   & 0.47 & 1132 \\
 6 &  98v &  500 &  7.0 & 1093 & 0.45-0.85 & 0.68-0.56 &  7.8-10.3 & 1.5-1.9   & 0.64 &  829 \\
 7 & 102j &  525 &  7.0 & 1094 & 0.43-0.85 & 0.79-0.79 &  8.0-9.6  & 1.5-1.6   & 0.87 & 1207 \\
 8 &  60g &  700 &  7.0 & 1091 & 0.45-0.85 & 0.89-0.89 &  8.4-10.5 & 1.6-2.0   & 1.58 & 1159 \\
 9 & 101k &  200 & 15.0 & 504  & 0.45-0.85 & 1.0-1.0   &  8.5-10.1 & 3.5-4.1   & 0.18 &  712 \\
10 & 119j &  500 & 45.0 & 1358 & 0.45-0.85 & $-$       &  8.2-9.5  & 1.2-1.4   & $-$  & 3359 \\
\noalign{\smallskip} \hline \noalign{\smallskip} 
\end{tabular} 
Telescopes brief descriptions:\\
1. Prime focus corrector. \\ 
2. Corrected Cassegrain. \\
3. Schmidt with a 3-lens corrector-flattener. \\
4. Schmidt (1934) - Houghton (1942, 1944). \\
5. Modified Richter-Slevogt. \\
6. Richter-Slevogt with a Mangin-type primary (Amon et al. 1971). \\ 
7. Double-pass singlet corrector with a Mangin primary ($\Omega_2$, Terebizh 2007a). \\
8. Double-pass 2-lens corrector with a Mangin primary ($\Omega_3$, Terebizh 2007a). \\
9. Refractive lens. \\
10. All-spherical system with a 4-lens corrector (Terebizh 2015). \\ 
}
\end{table}

General characteristics of the flat-field telescopes are given in Table~1, namely: 
the sequence number; number according to the 
author's catalogue; entrance pupil diameter $D$ (mm); angular diameter of the 
field of view $2w$ (degree); effective focal length $F$ (mm); spectral range used 
in calculations ($\mu$m); fraction of unvignetted rays $U$ on the optical axis 
and on the edge of a field; diameter $D_{80}$ of a circle that contains 80\% of 
the energy in a polychromatic star's image ($\mu$m and arc seconds); sky survey 
rate $\Gamma$ ({\it Herschels}), and the total length of the system from the 
first optical surface to the detector $L$~(mm). The focal ratios are presented 
below in Table~2. 

The term `sky survey rate' and the corresponding measurement unit {\it `Herschel'} 
are explained in Section~5. All systems have been optimized in the integral light 
within the waveband boundaries specified in Table~1, but they can be used in a 
wider spectral range. The common linear obscuration coefficient $\eta$ is related 
to the parameter $U$ as $U=1-\eta^2$, both depending on the field angle. The 
effective aperture diameter $D_e = DU^{1/2}$. The image diameter $D_{80}$ 
corresponds to the waveband specified in Table~1; it is obvious that by using of 
narrow-band filters one can improve image quality. 

We do not consider some attractive systems providing good images in the field 
up to $3.5^\circ$, in particular, the three-mirror anastigmat by Dietrich Korsch 
(1972, 1977) and the Mersenne-Schmidt telescope by Maurice Paul (1935; see 
Willstrop 1984 as well). The reason is that for the aperture less than 1~m one can 
attain the same image quality and field size with simpler optics and lower 
obscuration. In contrast, in the case of large telescopes, where the full-aperture 
field correctors cannot be applied, the systems mentioned above provide maximal 
depth of the survey. Thus, the layouts by Korsch and Paul were selected, respectively, 
for the space telescope SNAP of $2$~m aperture with the field of $1.5^\circ$ diameter 
and for the $8.4$~m ground-based telescope LSST with the field $3.5^\circ$.

Full descriptions of all the systems are given in the Appendix to this paper.

\begin{figure}   
   \centering
   \includegraphics[width=85mm]{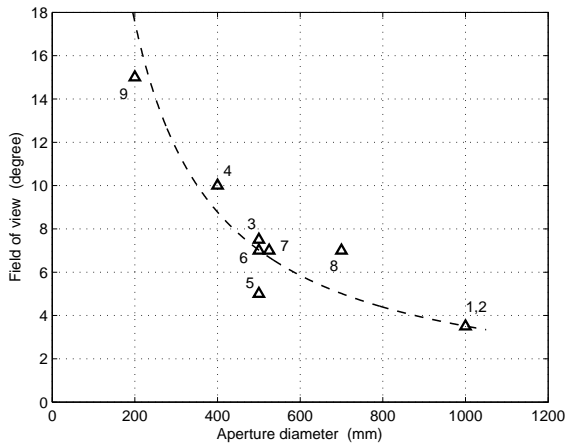}
   \caption{Relation between the diameters $D$ (mm) of flat-field systems in 
   Table~1 and their angular field of view 2$w$ (deg). 
   Dashed line corresponds to the approximate equation $2w = 3510/D$.}
\end{figure} 

Fig.~2 gives an idea of the telescope aperture $D$ and the angular field of view 
diameter $2w$ for systems listed in Table~1. As in the much more extensive set 
of existing telescopes discussed by Terebizh (2011), the field of view size is, 
to a first approximation, inversely proportional to the aperture diameter: 
$2w^\circ \simeq 3.51/D_m$. This dependence arises from a simple relation
\begin{equation}
  B \simeq D\phi \cdot 2w^\circ/57.3,
\end{equation}
given that the linear size of the detector $B$ and focal ratio $\phi \equiv F/D$ 
are changing insignificantly. In our case, this condition is satisfied, because 
the detector diagonal is fixed and $\phi$ are close to the mean value 
$\langle \phi \rangle \simeq 2.20$ (see Table~2).

Since aberrations of optical systems grow rapidly with increasing of their 
speed, a small spread in $\phi$ is specific for wide-field telescopes. Thus, 
the relation (1) clearly shows the need to increase the size of the detectors
for large survey telescopes. 

The primary mirrors of the two first telescopes from Table~1 have aspheric 
surfaces. The main problem that arises when manufacturing such systems is not 
as much concerned with the maximum deviation of the mirror's surface from 
sphere but with the {\it asphericity gradient} $G$ ($\mu$m/mm), i.e., the rate 
at which the deviation changes along the radial coordinate. An approximate 
expression for the maximum asphericity gradient $G_{max}$ of a conic section 
as a function of its diameter $D$, curvature radius at vertex $R_0$ and 
eccentricity $\varepsilon$ is: 
\begin{equation}
  G_{max} \simeq 31.25\, \varepsilon^2 (D/|R_0|)^3,\quad \mu m/mm
\end{equation}
(Terebizh 2011). Usually, the asphericity gradient of the primary mirror 
does not exceed $0.6\,\mu$m/mm. For the fastest existing wide-field telescopes
this value reaches $1.5\,\mu$m/mm (see Fig.~4 in the paper mentioned above). 
Here, the values of $G_{max}$ are $0.45\,\mu$m/mm and $0.58\,\mu$m/mm for 
the systems No.~1 and No.~2, respectively, i.e. they are relatively small. 

Although a corrected Cassegrain system is, in some essential respects, inferior 
to a system with the prime-focus corrector, its compactness may play the 
decisive role if necessary to make a number of identical instruments. 

In this regard, it is worth adding that the image quality in corrected Cassegrain 
systems depends weakly on the shape of the secondary mirror. In those cases, 
where the squared eccentricity of this mirror reaches values of the order of 20 
or even higher (for example, in PAN-STARRS; Hodapp et al. 2004), the gain in the 
image quality is totally smeared out by the manufacturing problems and severe 
tolerances in operation. On the contrary, tolerances are much more loose for 
systems with spherical secondary mirrors, which significantly increases the 
productivity of observations (an obvious example is GEODSS; Jeas 1981). The 
reason for such a small influence of the secondary mirror's shape in wide-field 
telescopes is clear: this mirror should be equally optimal for the light beams 
falling on it at very different angles, and this is possible only when the 
secondary mirror is close to a sphere.

\subsection{Spherical focal surface}

The general description of an all-spherical telescope with extremely wide field 
of view and spherical focal surface was given by Terebizh (2015, 2016). A few 
examples and the corresponding discussion were given in those papers. 
We propose here one more example of an $f/2.7$ system with an aperture of 500~mm   
and a $45^\circ$ angular field of view. It is included in Table~1 at No.~10; see 
also Fig.~1 and Fig.~3 for the optical layout and spot diagram; Appendix includes 
a complete set of parameters. Since the parameters $U$ and $\Gamma$ depend on the 
shape of detector, their values are left undefined. Generally speaking, the lenses 
can be made of arbitrary types of glass; in this case we choose the fused silica. 
In spite of a huge field size, the system provides good images: the $D_{80}$ image 
diameter in the polychromatic waveband $0.45-0.85\,\mu$m varies across the field 
in the range $1.2''-1.4''$, whereas the Airy disc diameter is $0.64''$. 

Evidently, the radical expanding of angular field of high-quality images is 
a consequence of the transition to a purely spherical optics. Just this feature 
provides the real \textit{point symmetry} of the system that is limited only by 
inevitable vignetting on the aperture stop. The second feature of the system that 
provides almost complete absence of chromaticity, is the afocality of the 4-lens 
corrector: it works at a $f/87$ speed. Finally, the third property necessary for 
ensuring wide field is the close proximity of the entrance pupil to the aperture 
diaphragm; their separation is only 31~mm in a case under consideration. 

\begin{figure}  
   \centering
   \includegraphics[width=85mm]{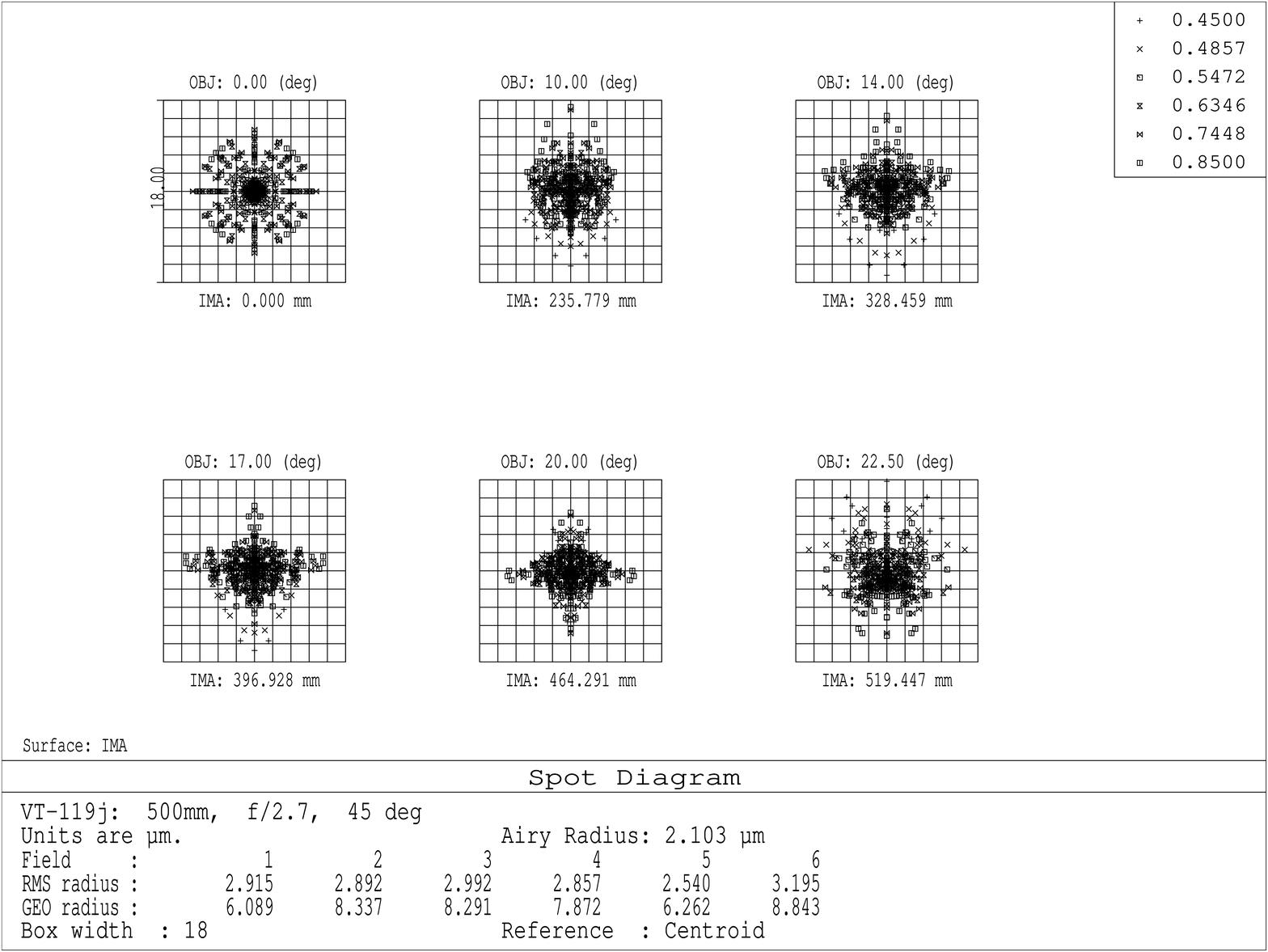}
    \caption{Spot diagram of VT-119j design in the polychromatic waveband 
	 $0.45-0.85\,\mu$m. The field angles are $0^\circ$, $10^\circ$, $14^\circ$, 
	 $17^\circ$, $20^\circ$ and $22.5^\circ$. Airy disc diameter is $4.2\,\mu$m, 
	 box width is $18\,\mu$m $\simeq 2.7''$.}
\end{figure}

As mentioned in the Introduction, the main problem with such systems is the need 
to work with the curved focal surface. The following ways seem to be preferred 
now in this regard: \textit{i})~the use of large detectors with a curved surface; 
\textit{ii})~applying the long-known technology based on a plurality of delicate 
waveguides with a curved in aggregate input faced to the focal surface;
\textit{iii}) using a small flat detectors equipped with a flattening optics.

The principal issues and examples of curved detectors were discussed by Iwert 
and Delabre (2010), Iwert et al. (2012); the first of these papers includes a 
photograph of a curved detector with size of 60~mm $\times$ 60~mm and curvature 
radius 500~mm. There are also working devices of this type. In particular, curved 
detector has been implemented in the DARPA 3.5~m Space Surveillance Telescope 
(Blake et al. 2013). It is important to note that individual detectors of any 
form can be placed, both continuously and discretely, on the curved focal surface 
according to the shape of the studied area of the sky.

The second option, being considered in a modern context, involves a number of 
technological problems. One should expect that basically these problems will be 
solved within the framework of the announced by European Space Agency in 2013 
program, which provides a solution for mapping a curved image field onto a flat 
imaging detector array.

We mentioned the third option in Sec.~1 in connection with the \textit{Kepler} 
telescope. In a case of VT-119j, the curvature radius of the focal surface is 
rather large, 1329~mm, so an additional flattening optics can be made of a single 
lens. Let us consider, for example, a flat detector of format 30~mm $\times$ 
30~mm, i.e., with a diagonal length 42.4~mm. If we keep the quality of central 
images, then their corner diameter will be about 50~$\mu$m. We can made the blur 
of spots across the field less than 23~$\mu$m by shifting the detector at 0.1~mm. 
Finally, the image quality is restored completely, when a weak lens made 
of fused silica is installed in front of the detector. Since the lens radii of 
curvature are of about 300~mm and 200~mm, one can simply use it as the detector 
window. 

Thus, in the case under consideration, the third option is the most 
straightforward.

\section{Sampling factor}

As one can see, the value of $D_{80}$ characterizes the image quality provided 
by a telescope alone; to distinguish $D_{80}$ from similar quantities we 
designate it hereafter as $\beta_{tel}$. For the telescopes considered in this 
paper, it varies from $0.6''$ up to $2.9''$, i.e., has the same order of 
magnitude as the atmospheric blurring $\beta_{atm}$. For our purposes, it is 
enough to accept that the angular diameter of a star image due to these two 
factors is
\begin{equation} 
  \beta = \sqrt{\beta_{atm}^2 + \beta_{tel}^2}.
\end{equation} 
Accepting constant $\beta_{atm} = 1.5''$ and $\beta_{tel}$ according to Table~1,  
we obtain the resulting values of image quality $\beta$ shown in the fourth 
column of Table~2. Let us accept also the linear size of the detector's pixel 
equal to $9\,\mu$m, as in the CCD STA 1600. The corresponding angular sizes of 
pixels are given in the fifth column of Table~2.   

Among the set of parameters that define the area of application of a telescope, 
an important role plays the {\it sampling factor} $\chi$~-- the ratio of the 
diameter of the star image to the size of the pixel:
\begin{equation}
  \chi \equiv \beta/p, 
\end{equation}
where both $\beta$ and $p$ are either angular or linear. According to the 
well-known {\it sampling theorem} by V.~Kotel'nikov and C.~Shannon (see, e.g., 
Press et al. 1992, p.~500), the retaining of the entire spectrum of spatial 
frequencies of a continuous image at sampling requires the discretization step, 
which not exceeds $\delta x_c \equiv 1/(2f_c)$, where $f_c$ is the {\it cutting 
frequency} above which the power spectrum of the continuous object can be 
considered negligible. In typical astronomical applications $\delta x_c$ is 
approximately equal to half a radius of the star's image. Thus, one usually 
should have at least $4$ pixels covering the diameter of a star image (i.e. 
$\chi \simeq 4$). Taking into account random fluctuations of flux, this value 
is usually increased to $8$ in precise photometric measurements ($\chi \simeq 7$ 
for the {\it Kepler} space telescope). On the other hand, in surveys, where 
detecting faint objects is of the primary importance, the sampling factor is 
reduced to $1-2$. 

The $\chi$ values corresponding to the conditions we have adopted are given 
in the last column of Table~2. We can see that the discussed sample of 
telescopes exactly suites the work of search or exploratory nature.

\begin{table}   
\centering 
\caption{Focal ratio $\phi \equiv F/D$ and sampling factor $\chi$ at the 
atmosphere  blurring $1.5''$ and linear pixel size $9\,\mu$m.} 
\begin{tabular}{cccccc}
\hline \noalign{\smallskip} 
 System & $\phi$ & Scale   & $\beta''$ & $p''$  & $\chi$\\
 No.    &        & $\mu$m/$''$ &       &        &       \\
\hline
  1     & 2.18   & 10.6    & 1.62      & 0.85   & 1.9 \\
  2     & 2.19   & 10.6    & 1.83      & 0.85   & 2.2 \\
  3     & 2.05   & 4.97    & 2.42      & 1.81   & 1.3 \\
  4     & 1.91   & 3.71    & 3.05      & 2.43   & 1.3 \\
  5     & 3.08   & 7.46    & 2.16      & 1.21   & 1.8 \\
  6     & 2.19   & 5.30    & 2.27      & 1.70   & 1.3 \\
  7     & 2.08   & 5.30    & 2.16      & 1.70   & 1.3 \\
  8     & 1.56   & 5.29    & 2.34      & 1.70   & 1.4 \\
 10     & 2.72   & 6.58    & 1.98      & 1.37   & 1.4 \\
\noalign{\smallskip} \hline
\end{tabular} 
\end{table}

\section{Limiting magnitude and survey speed}

Let us now consider the characteristics of wide-field telescopes that are of 
special interest within the scope of this paper. Namely, these are the limiting 
magnitude ($m_{lim}$) and survey speed ($S$, deg$^2$/sec) determined by the 
telescope\,+\,detector system and observational conditions. In calculations, we 
took into account the entrance pupil diameter, effective focal length, 
angular field of view, telescope transparency, fraction of unvignetted rays, 
bandwidth, and $\beta_{tel}$ value. Parameters of the detector are the same for 
telescopes: the quantum efficiency is $0.85$~events/photon and the pixel size is 
$9\,\mu$m (CCD STA 1600). It was assumed that the noise obeys the Poisson 
distribution. As concerns the observational conditions, we have assumed that 
$\beta_{atm} = 1.5''$, sky background is $20.0^m$/arcsec$^2$, optical thickness 
of the atmosphere in zenith is $0.30$, object zenith angle is $40^\circ$, dead 
time is $5$~sec, and the threshold signal-to-noise ratio $S/N = 8$. The parameter 
we call `dead time' is the sum of read-out and telescope redirection time spans. 
The $S/N$ ratio corresponds to the total number of pixels in detector. The variable 
value in our calculations is the exposure time~$T$. We tried to adopt the above 
parameters as close as possible to their typical values. Of course, variations 
of initial values change the results, but not radically. 

Figures~4 and 5 demonstrate resulting values of the limiting magnitude and 
survey speed, respectively. The first are in a quite good agreement with the 
estimates according to the SIGNAL package created by the team of Isaac 
Newton Group of Telescopes (http://catserver.ing.iac.es/signal/). 
However, we do not require the calculations to exactly match the real data, 
because our primary goal is to evaluate the comparative characteristics of
various types of optical systems.

\begin{figure}  
   \centering
   \includegraphics[width=85mm]{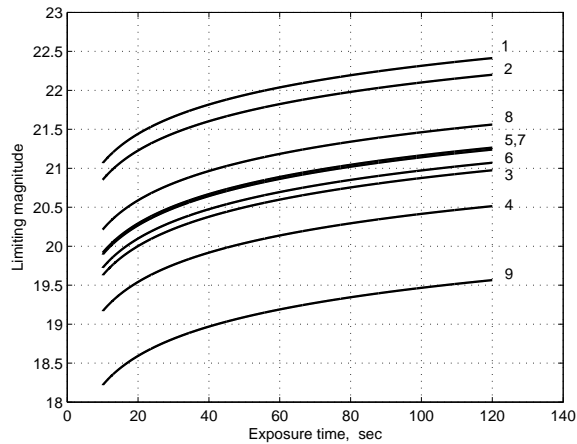}
   \caption{Limiting magnitude as a function of the exposure time 
     for the flat-field systems in Table~1.}
\end{figure}

\begin{figure}   
  \centering
  \includegraphics[width=85mm]{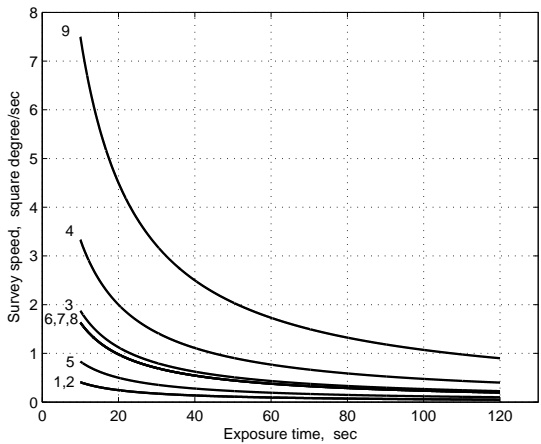} 
  \caption{Survey speed as a function of the exposure time 
  for the flat-field systems in Table~1.} 
\end{figure}

As one can see, flat-field optical systems are clearly divided into three 
groups: {\it i}) systems No.~1 and No.~2; {\it ii}) systems No.~$3-8$; {\it iii}) 
lens objective No.~$9$. This division results from the initial grouping  
according to $D$ and $2w$ values (see Fig.~2), because the limiting magnitude  
depends primarily on the aperture diameter and does not depend on the field size,
while the survey speed is proportional to the field area, which is maximal for 
relatively small telescopes.

Since $m_{lim}$ grows and $S$ decreases with increasing of $T$, we can expect 
that some their combination is independent, to a first approximation, on the 
exposure time. Indeed, the limiting magnitude $m_{lim} = 2.5 \log(DT^{1/2}) + A$,
where $A$ incorporates all other parameters. Taking into accound that, for 
relatively short dead time, the survey speed is $S \simeq (2w)^2/T$, we obtain:
\begin{equation}
  Q \equiv m_{lim} + 1.25 \log S \simeq const.
\end{equation}
In our case, at $T$ in the range from 10~sec to 90~sec, the RMS variation of 
$Q$ for any particular design is only $0.04$. The individual values of $Q$ vary  
from 19.5 for the 200-mm refractor to 20.7 for the 1-m system with a prime-focus 
corrector. Obviously, the $Q$ value depends not only on the telescope\,+\,detector 
system, but also on the observational conditions, so, the bigger $Q$ we reach, 
the more effective observational system we have. An adequate criterion of 
effectiveness of the telescope alone is discussed below in Section~5.

When planning a program of observations, different priority is assigned usually
to the achieved stellar magnitude and survey speed. Therefore, the main problem 
is choosing an appropriate exposure time $T$ required to obtain the desired 
$m_{lim}$ and $S$ values. In turn, the adopted by an observer triplet 
$\{m_{lim},S,T\}$ determines the proper choice of the telescope, namely, its 
optical system and the set of specifications listed in Table~1.

Let us suppose, for example, that we should provide $S \sim 1$~deg$^2$/sec,
as mentioned in the Introduction, and the limiting magnitude $m_{lim} \simeq 20.5$.
Fig.~4 shows that telescopes No.~1 and No.~2 are too big for this purpose, because 
the needed exposure time is short in comparison with the dead time. The required 
limiting magnitude can be achieved by using systems No.~3, 5 -- 8 at quite an
appropriate exposure time of $25-50$~sec. However, the further consideration of 
Fig.~5 excludes the system No.~5 (modified Richter-Slevogt), because it provides 
too small field. The system No.~3 (Schmidt camera) contains the corrector with a 
complicated aspheric surface; besides, to achieve good image quality one should 
use an expensive and not simple in making glass. These considerations impel us to 
opt for one of the all-spherical telescopes with simple types of glass No.~6 
(Amon et al.), No.~7 ($\Omega_2$) or No.~8 ($\Omega_3$). The final choice 
requires the detailed discussion of more subtle properties of these systems. 

To show the essential conditionality of choosing a proper optical layout, we note 
that system No.~4 (all-spherical Schmidt-Houghton with a three-lens input 
corrector) is clearly preferable at necessity to achieve the same survey speed of 
$S \sim 1$ deg$^2$/sec but $m_{lim} \sim 20.0$ for about 40~sec exposure. 
A number of modified telescopes of this type were made within the last decade 
(Terebizh 2011).

Naturally, only 1-meter telescopes No.~1 and No.~2 allow detecting the faint 
objects in the range of $21.5^m - 22.5^m$. The difference in the limiting 
magnitude between these two systems is $\sim0.21^m$; it equally results from a 
lower obscuration and better image quality provided by the prime-focus corrector. 
Both of these factors are typical for the compared systems; the installing 
of the secondary mirror only allows achieving compactness of the telescope at a 
predetermined effective focal length.

We do not include the effect of stray light and direct exposure of detector by
sky background, because these factors are highly dependent not only on the 
optical scheme of the telescope, but also on its structure. Evidently, in this 
respect the telescope with a prime-focus corrector again clearly superior to 
systems such as the corrected Cassegrain. In practice, the need to install a 
complicated system of baffles in a wide-field Cassegrain design reduces both 
limiting magnitude and size of the field.

If necessary, the limiting magnitude of a survey telescope can be somewhat 
increased by reducing its angular field of view but leaving unchanged the linear 
dimensions of the detector. This increases the focal length of the telescope, 
and, as a consequence, reduces aberrations, improves image quality and reduces 
the contribution of the sky background in a pixel of smaller angular size.

As regards the design No.~10 with a field of $45^\circ$ in diameter, the 
corresponding limiting magnitude and survey speed depend on the relative area of 
detectors in the focal surface. Let us suppose, for example, that a strip of 
sizes $45^\circ \times 4^\circ$ ($\sim 1000$~mm $\times$~95~mm) is occupied by 
detectors. Then, the fraction of unvignetted rays $U \simeq 0.89$, which gives  
the equivalent aperture diameter $D_e \simeq 472$~mm. Thus, the limiting 
magnitude for the system No.~10 is approximately equal to those for the system 
No.~7 (see Fig.~4). 

Note that the strip's area $180$\,deg$^2$ corresponds to diameter of the 
equivalent circular field of view $2w_e \simeq 15.1^\circ$, so the expected 
survey speed is very high. In particular, at the exposure time 20~sec and dead 
time 5~sec we obtain $S \simeq 7$~deg$^2$/sec; at such a speed, the sky region 
of area $10^4$\,deg$^2$ will be examined in less than half an hour.

\section{Sky survey rate}

Along with a number of standard parameters of telescopes, it would be useful 
to have a parameter, giving an idea of the efficiency of the telescope just as 
a survey tool. To date, such a parameter widely used is the {\it etendue} 
$E \equiv \pi w^2\cdot \pi D_e^2/4$, a product of the observed sky area (deg$^2$)
and the effective area of the telescope aperture (m$^2$). The inadequacy of 
this measure is evident from the fact that $E$ does not take into account the 
quality of images provided by survey telescope. However, there is no doubt that
the better angular resolution we can reach, the higher is the efficiency of the
survey. An adequate measure of survey efficiency, the {\it sky survey rate} 
$\Gamma$, was proposed by V.V.~Biryukov and the author (Terebizh 2011). In the 
case when the dead time is much less than the exposure time, the sky survey rate 
is proportional to the ratio of the observed sky area $\pi w^2$ to the exposure 
time, needed to achieve the required $S/N$ value. It is not difficult to show 
that parameter defined in this way is
\begin{equation}
  \Gamma \equiv \frac{\pi w^2 \cdot \pi D_e^2/4  }{\Delta^2} = E/\Delta^2,
\end{equation}
where $\Delta$ is the so called {\it delivered image quality}, measured 
in angular units. For our purpose, it is sufficient to accept  
\begin{equation}
  \Delta = \sqrt{\beta_{atm}^2 + \beta_{tel}^2 + (p'')^2}.
\end{equation}
In essence, $\Gamma$ is the product of the number of resolution elements 
in the observed region of the sky and the effective area of the telescope's 
aperture\footnote{J.~Tonry~(2010) proposed a more detailed approach to the 
evaluation of the survey effectiveness, which includes consideration of the 
Point Spread Function form and its alignment with the pixels of the detector.}.
For practical needs a convenient measurement unit of $\Gamma$ is
\begin{equation}
  Herschel \equiv 1\,m^2 deg^2/arcsec^2,
\end{equation}
named after William Herschel (1738-1822). Hereafter we use an abbreviation $H$.

Values of $\Gamma$ for the systems considered here are given in Table~1. 
First of all, pay attention to the marked superiority of the telescope No.~1 on 
the system No.~2, which shows once again the merits of the location the lens 
corrector in a prime focus. 

Then, one might suppose that system No.~4 is significantly more effective than 
system No.~5, since their apertures do not differ very much, but the field of 
view of system No.~4 is twice as larger. However, in reality they have the same 
values of $\Gamma$, which stresses the importance of good image quality for 
survey telescopes.

As for the lens objective No.~9, its small diameter and a comparatively lower 
quality of images do not provide large value of $\Gamma$. Nevertheless, very 
high survey speed makes it quite suitable for observations of rapidly varying 
objects. For example, at the exposure time 20~sec the survey speed reaches 
$4.5$~deg$^2$/sec, which allows registering of objects brighter than $18.6^m$ 
in the sky area of $10^4$ deg$^2$ in just 40 minutes. If necessary, the image 
quality of the lens can be improved by a slight aspherization of certain 
surfaces or by using a special glass. 

The high efficiency of the $\Omega_{2-3}$-designs No.~7 and No.~8 
draws attention. These all-spherical systems offer good images in a large field 
of view; the secondary passage of light through the input optics enables 
significant reducing both the aberrations and light obscuration, so the 
effective aperture diameter is not much inferior to the entrance pupil diameter. 
As is known, the use of spherical optics significantly mitigates the tolerances  
in the manufacture and operation of telescopes; ultimately it strongly affects 
the total cost of the sky survey systems. It can be assumed that these systems 
will be widely used in future surveys.

As one might expect, the design No.~10 provides the highest survey rate. 
Continuing discussion in a frame of the example of preceding section, we have 
the field area $180$\,deg$^2$, the equivalent aperture diameter $D_e = 0.472$~m 
and the delivered image quality $\Delta = 2.41''$. According to equation (6),
the resulting survey rate $\Gamma \simeq 5.4$~{\it H}, which is almost 3 times
greater than the maximum value for the systems in Table~1.

\section{Concluding remarks} 

Since we are primarily interested in the {\em performance} of various optical 
systems, we touch only in passing the issues relating to their manufacturing. 
For various reasons, the technological challenges are growing rapidly with 
increasing the speed of a system, say, when the focal ratio becomes less than 
$1.5$. As Table~2 shows, this is not the case for the sample of designs under 
consideration. 

It is also worth adding that each of the designs No.~1-3 contains only one 
aspheric surface. Thus, the tolerances are quite reasonable for these systems, 
while for the all-spherical designs No.~4-10 the tolerances are so mild as 
possible given a system's speed.

We have seen that even a single wide-field telescope with an aperture less than
1~m and a flat detector, designed and manufactured properly, ensure the 
registration of objects brighter than $21.5^m - 22.5^m$ in the entire visible 
hemisphere of the sky within one night. Much faster, but not so deep survey 
provides a system with the spherical focal surface. Of course, the more extensive 
and reliable data should come from a hierarchic observational set, comprised of 
a few telescopes of different types with optimally chosen specifications. Such 
systems, using previously made telescopes work now effectively, in particular, 
the Palomar Transient Factory (Law, Kulkarni et al. 2009) and Catalina Real-Time 
Transient Survey (Djorgovski, Drake et al. 2011). The ATLAS (Asteroid 
Terrestrial-impact Last Alert System, Tonry 2010), using newly built instruments, 
is close to completion. 

As far as the choice of a telescope's optical scheme substantially depends on 
a specific task of wide-field observations, there is no preferred option 
suitable for all occasions. We also need to add the well-known fact that the 
choice of an optical layout largely depends on the characteristics of the 
assumed detector of light. However, the two mentioned factors~-- problem 
to be solved and the properties of detector, which, of course, one need to add 
an assessment of the project's cost~-- almost uniquely define the required 
optical scheme.

\section*{Acknowledgments}

I thank M.R.~Ackermann (University of New Mexico, U.S.A.), 
V.V.~Biryukov (Crimean Astrophysical Observatory), 
M.~Boer (Recherche CNRS ARTEMIS, France),
D.A.~Kononov (Institute of Astronomy, Russian Academy of Sciences), 
Yu.A.~Petrunin (Telescope Engineering Company, U.S.A.) and 
J.L. Tonry (Institute for Astronomy, University of Hawaii, U.S.A.) 
for useful comments to the paper.

\section*{Appendix. Description of optical systems}

The following is a complete description of the designs, which are included 
in Table~1 at Nos.~1~- 8 and No.~10. Numerical data are presented in a common 
format in the Tables 3~- 11, the corresponding optical layouts are shown in 
Fig.~1. 

The numbering of surfaces in tables corresponds to the optical path of light. 
The terms `aperture' and 'entrance pupil' are considered as equivalent. All 
distances are given in millimeters. The curvature radii and inter-element 
distances are presented with the precision required by optical soft packages. 
We do not round the thicknesses of lenses, because variations of the optical 
constants of each type of glass at manufacturing will inevitably require 
minor adjustment of thicknesses. 

The sorts of optical glass correspond to those of catalogues by Schott 
(N-\ldots) and Ohara (S-\ldots). In both catalogues, the selection of
glass was limited by the maximum melt frequency. Sorts S-BSL7 and N-BK7 are
equivalent. The data source for Fused silica is `The Infrared and 
Electro-Optical Systems Handbook', Vol.~III, Ch.~1. 

Since the filter and detector window have zero optical power, their position
and thickness can be changed with slight correction of the basic optical
layout. 

As usual, optical schemes can be scaled up or down. Since image quality is not 
far from the diffraction limit, the scaling up should be performed along with 
slight optimization.

The brief designations are the following: $R_0$~-- paraxial curvature radius, 
$T$~-- distance to the next surface, $D$~-- light diameter, FS~-- fused silica, 
Stop~-- aperture stop, SP~-- stop position, L$_k$~-- k-th lens, Pri~-- 
primary mirror, Sec~-- secondary mirror, Obsc~-- obscuration, F~-- filter, 
W~-- window of the detector, Ima~-- image on the focal surface.

\begin{table}[h]   
\caption{VT-56y design with an aperture of 1.0~m and $3.5^\circ$ 
field. The effective focal length is 2183~mm. 
The design waveband is $0.40-0.85 \mu$m.}
\begin{tabular}{cccccc}
\hline \noalign{\smallskip} 
Surf.& Com-       & $R_0$     & $T$       & Glass   & $D$    \\
 No. & ments      & (mm)      & (mm)      &         & (mm)   \\
\hline
 1   & Stop       & $\infty$  & 26.509    & --      & 1000.0 \\
 2   & Pri        & -4708.58  & -1645.73  & Mirror  & 1000.0 \\
 3   & L$_1$      & -419.235  & -45.0     & N-BK7   & 391.6  \\
 4   &            & -1199.06  & -101.877  & --      & 384.5  \\
 5   & L$_2$      & -504.582  & -24.40    & FS      & 304.5  \\
 6   &            & -244.457  & -61.976   & --      & 274.4  \\
 7   & L$_3$      & 1854.23   & -34.50    & N-LAK8  & 267.9  \\
 8   &            & -344.778  & -23.148   & --      & 255.9  \\
 9   & L$_4$      & -2822.80  & -20.0     & S-PHM52 & 256.2  \\
10   &            & 757.208   & -187.357  & --      & 256.9  \\
11   & L$_5$      & -406.165  & -42.40    & S-FPL53 & 252.7  \\
12   &            & 496.992   & -148.232  & --      & 251.4  \\
13   & F          & $\infty$  & -7.0      & N-BK7   & 157.6  \\
14   &            & $\infty$  & -18.0     & --      & 155.1  \\
15   & W          & $\infty$  & -6.0      & FS      & 144.8  \\
16   &            & $\infty$  & -14.0     & --      & 142.5  \\
17   & Ima        & $\infty$  & --        & --      & 134.5  \\
\noalign{\smallskip} \hline
\end{tabular}

\vspace*{5pt}
{\small Notes to Table~3:\\ 
Conic constant of the primary mirror is $-$1.506709, all other 
surfaces are spheres. \\
Light obscuration corresponds to the round screen of diameter 392~mm.}
\end{table}

\begin{table}[h]   
\caption{VT-112m design with an aperture of 1.0~m and $3.5^\circ$ 
field. The effective focal length is 2189~mm.
The design waveband is $0.45-0.85 \mu$m.} 
\begin{tabular}{cccccc}
\hline \noalign{\smallskip}
Surf.& Com-    & $R_0$     & $T$      & Glass   & $D$    \\
 No. & ments   & (mm)      & (mm)     &         & (mm)   \\
\hline
 1   & Stop    & $\infty$  & 28.944   & --      & 1000.0 \\
 2   & Pri     & -4311.49  & -1139.21 & Mirror  & 1000.0 \\
 3   & Sec     & -11457.9  & 601.547  & Mirror  & 549.5  \\ 
 4   & L$_1$   & 342.969   & 32.0     & N-BK7   & 350.0  \\
 5   &         & 239.293   & 44.296   & --      & 322.7  \\
 6   & L$_2$   & 622.257   & 32.0     & N-LAK9  & 322.6  \\
 7   &         & 20122.9   & 130.451  & --      & 320.0  \\
 8   & L$_3$   & 766.109   & 26.0     & N-LAK10 & 265.7  \\
 9   &         & 249.604   & 102.599  & --      & 248.2  \\
10   & L$_4$   & 363.158   & 70.0     & S-FPL53 & 263.8  \\
11   &         & -294.997  & 94.073   & --      & 262.1  \\
12   & L$_5$   & -595.004  & 17.360   & N-LAK14 & 178.9  \\
13   &         & -11372.6  & 30.005   & --      & 172.6  \\
14   & F       & $\infty$  & 5.0      & N-BK7   & 158.6  \\
15   &         & $\infty$  & 25.0     & --      & 157.1  \\
16   & W       & $\infty$  & 4.0      & FS      & 145.5  \\
17   &         & $\infty$  & 21.0     & --      & 144.2  \\
18   & Ima     & $\infty$  & --       & --      & 134.5  \\
 \noalign{\smallskip} \hline
\end{tabular}
 
\vspace*{5pt}
{\small Notes to Table~4: \\  
Conic constant of the primary mirror is $-$1.495564, all other
surfaces are spheres. \\
Light obscuration corresponds to the round screen of diameter 560~mm.}
\end{table}

\begin{table}[h]   
\caption{VT-110f design with an aperture of 500~mm and $7.5^\circ$ 
field. The effective focal length is 1026~mm.
The design waveband is $0.42-0.82 \mu$m.} 
\begin{tabular}{cccccc}
\hline \noalign{\smallskip}
Surf.& Com-    & $R_0$     & $T$      & Glass   & $D$    \\
 No. & ments   & (mm)      & (mm)     &         & (mm)   \\
\hline
 1   & L$_1$   & $\infty$  & 25.142   & FS      & 502.2 \\
 2   & Stop    & $\infty$  & 1472.789 & --      & 500.0 \\
 3   & Pri     & -2336.261 & -979.429 & Mirror  & 680.2  \\ 
 4   & L$_1$   & -376.877  & -24.832  & SF10    & 225.0  \\
 5   &         & 9297.360  & -2.035   & --      & 221.3  \\
 6   & L$_2$   & 2346.990  & -18.356  & N-LAF7  & 221.2  \\
 7   &         & -210.890  & -3.828   & --      & 198.8  \\
 8   & L$_3$   & -196.733  & -45.723  & S-FPL53 & 198.2  \\
 9   &         & 946.597   & -40.557  & --      & 192.6  \\
10   & F       & $\infty$  & -4.0     & N-BK7   & 160.6  \\
11   &         & $\infty$  & -16.0    & --      & 158.8  \\
12   & W       & $\infty$  & -4.0     & FS      & 147.6  \\
13   &         & $\infty$  & -16.0    & --      & 145.7  \\
14   & Ima     & $\infty$  & --       & --      & 134.5  \\
\noalign{\smallskip} \hline
\end{tabular}
 
\vspace*{5pt}
{\small Notes to Table~5: \\ 
Obscuration on surface No.~2 of diameter 250.0~mm. \\
Surface No.~2 is even asphere with $A_2 = -2.08187\cdot 10^{-5}$, 
$A_4 = 4.811443\cdot 10^{-11}$. \\
All other surfaces are spheres.}
\end{table}

\begin{table}[h]   
\caption{VT-78e design with an aperture of 400~mm and $10^\circ$ 
field. The effective focal length is 764~mm.
The design waveband is $0.45-0.85 \mu$m.} 
\begin{tabular}{cccccc}
\hline \noalign{\smallskip}
Surf.& Com-    & $R_0$     & $T$      & Glass   & $D$    \\
 No. & ments   & (mm)      & (mm)     &         & (mm)   \\
\hline
 1   & L$_1$   & 1077.981  & 59.681   & N-BK7   & 484.8 \\
 2   &         & -1221.197 & 78.931   & --      & 482.6 \\
 3   & L$_2$   & -1251.780 & 37.591   & N-BK7   & 425.4  \\ 
 4   &         & 2367.109  & 39.113   & --      & 406.3  \\
 5   & L$_3$   & -694.206  & 39.247   & N-BK7   & 406.3  \\
 6   &         & -1770.605 & 262.603  & --      & 405.4  \\
 7   & Stop    & $\infty$  & 15.726   & --      & 365.3  \\
 8   & Pri     & -1072.895 & -278.328 & Mirror  & 365.3  \\
 9   & Sec     & -1770.605 & 279.256  & Mirror  & 239.0  \\
10   & L$_4$   & 398.439   & 40.0     & N-SK10  & 154.0  \\
11   & L$_5$   & -395.695  & 25.0     & P-LAF37 & 154.0  \\
12   &         & -36384.17 & 29.359   & --      & 154.0  \\
13   & F       & $\infty$  & 5.0      & N-BK7   & 154.7  \\
14   &         & $\infty$  & 25.0     & --      & 153.3  \\
15   & W       & $\infty$  & 5.0      & FS      & 142.5  \\
16   &         & $\infty$  & 15.0     & --      & 141.0  \\
17   & Ima     & $\infty$  & --       & --      & 134.5  \\
 \noalign{\smallskip} \hline
\end{tabular}

\vspace*{5pt} 
{\small Notes to Table~6: \\  
Obscuration on surface No.~6 of diameter 188.0~mm. \\ 
All surfaces are spheres.}
\end{table}

\begin{table}[h]   
\caption{VT-77i design with an aperture of 500~mm and $5.0^\circ$ 
field. The effective focal length is 1538~mm.
The design waveband is $0.45-0.85 \mu$m.} 
\begin{tabular}{cccccc}
\hline \noalign{\smallskip}
Surf.& Com-    & $R_0$     & $T$      & Glass   & $D$    \\
 No. & ments   & (mm)      & (mm)     &         & (mm)   \\
\hline 
 1   & L$_1$   & 2667.080  & 44.0     & N-BK7   & 527.0  \\ 
 2   &         & 14019.45  & 278.784  & --      & 522.8  \\
 3   & L$_2$   & -1843.524 & 40.0     & N-BK7   & 478.0  \\
 4   & Stop    & -4428.837 & 646.791  & --      & 476.9  \\
 5   & Pri     & -2430.967 & -646.791 & Mirror  & 534.7  \\
 6   & Sec     & -4428.837 & 621.791  & Mirror  & 310.1  \\
 7   & L$_3$   & 802.225   & 45.0     & N-KZFS4 & 177.3  \\
 8   & L$_4$   & 174.767   & 36.0     & N-LAK12 & 164.3  \\
 9   &         & 1702.637  & 41.971   & --      & 160.0  \\
10   & F       & $\infty$  & 5.0      & N-BK7   & 145.9  \\
11   &         & $\infty$  & 10.0     & --      & 144.8  \\
12   & W       & $\infty$  & 3.0      & FS      & 141.2  \\
13   &         & $\infty$  & 17.0     & --      & 140.5  \\
14   & Ima     & $\infty$  & --       & --      & 134.5  \\
 \noalign{\smallskip} \hline
\end{tabular}

\vspace*{5pt}  
{\small Notes to Table~7: \\  
Obscuration on surface No.~4 of diameter 244.0~mm. \\
All surfaces are spheres.}
\end{table}

\begin{table}[h]   
\caption{VT-98v design with an aperture of 500~mm and $7^\circ$ 
field. The effective focal length is 1093~mm.
The design waveband is $0.45-0.85 \mu$m.} 
\begin{tabular}{cccccc}
\hline \noalign{\smallskip}
Surf.& Com-    & $R_0$     & $T$      & Glass   & $D$    \\
 No. & ments   & (mm)      & (mm)     &         & (mm)   \\
\hline
 1   & L$_1$   & 1961.024  & 65.083   & FS      & 582.3 \\
 2   &         & -1654.074 & 60.033   & --      & 579.1 \\
 3   & L$_2$   & -1858.116 & 39.028   & FS      & 550.2  \\ 
 4   &         & 79341.23  & 405.393  & --      & 538.4  \\
 5   & Stop    & $\infty$  & 22.105   & --      & 421.8  \\
 6   & L$_3$   & -1086.270 & 45.923   & FS      & 421.8  \\
 7   & Pri     & -2128.737 & -45.923  & Mirror  & 429.1  \\
 8   & L$_3$   & -1086.270 & -427.499 & --      & 414.0  \\
 9   & Sec     & 79341.23  & 515.142  & Mirror  & 323.1  \\
10   & L$_4$   & 305.528   & 23.460   & FK3     & 200.0 \\
11   &         & -3053.207 & 30.473   & --      & 198.2 \\
12   & L$_5$   & -661.789  & 16.418   & S-TIM22 & 182.7  \\
13   &         & -768.247  & 28.810   & --      & 178.7 \\
14   & F       & $\infty$  & 7.0      & N-BK7   & 159.5  \\
15   &         & $\infty$  & 23.0     & --      & 156.9  \\
16   & W       & $\infty$  & 10.0     & FS      & 143.9  \\
17   &         & $\infty$  & 10.0     & --      & 140.2  \\
18   & Ima     & $\infty$  & --       & --      & 134.5  \\
 \noalign{\smallskip} \hline
\end{tabular}

\vspace*{5pt} 
{\small Notes to Table~8: \\  
Obscuration on surface No.~4 of diameter 270.0~mm. \\
All surfaces are spheres.}
\end{table}

\begin{table}[h]   
\caption{VT-102j design with an aperture of 525~mm and $7.0^\circ$ 
field. The effective focal length is 1094~mm.
The design waveband is $0.43-0.85 \mu$m.} 
\begin{tabular}{cccccc} 
\hline \noalign{\smallskip}
Surf.& Com-    & $R_0$  & $T$       & Glass   & $D$    \\
 No. & ments   & (mm)   & (mm)      &         & (mm)   \\
\hline
 1 & Stop  & $\infty$   & 150.0     & --      & 525.0 \\
 2 & Obsc  & $\infty$   & 64.192    & --      & 240.0 \\
 3 & L$_1$ & 4516.956   & 61.791    & FS      & 552.3 \\
 4 &       & -4110.190  & 828.460   & --      & 554.3 \\
 5 & L$_2$ & -1249.675  & 53.0      & FS      & 558.6 \\
 6 & Pri   & -2262.803  & -53.0     & Mirror  & 568.4 \\
 7 & L$_2$ & -1249.675  &-828.460   & --      & 544.8 \\
 8 & L$_1$ & -4110.190  &-61.791    & FS      & 285.2 \\
 9 &       & 4516.956   &-64.192    & --      & 271.5 \\
10 & L$_3$ & -419.927   &-50.0      & S-PHM52 & 239.4 \\
11 &       & 1169.041   &-0.20      & --      & 226.8 \\
12 & L$_4$ & 1167.046   &-44.897    & S-LAH55 & 226.6 \\
13 &       &-16723.57   &-63.966    & --      & 207.4 \\
14 & F     & $\infty$   &-5.0       & S-BSL7  & 161.4 \\
15 &       & $\infty$   &-15.0      & --      & 159.1 \\
16 & W     & $\infty$   &-3.0       & FS      & 148.3 \\
17 &       & $\infty$   &-17.0      & --      & 146.8 \\
18 & Ima   & $\infty$   &           & --      & 134.5 \\
\noalign{\smallskip} \hline
\end{tabular}
 
\vspace*{5pt}
{\small Notes to Table~9: \\ 
Obscuration on surface No.~2 of diameter 240.0~mm. \\
All surfaces are spheres.}
\end{table}

\begin{table}[h]   
\caption{VT-60g design with an aperture of 700~mm and $7.0^\circ$ 
field. The effective focal length is 1091~mm.
The design waveband is $0.45-0.85 \mu$m.} 
\begin{tabular}{cccccc} 
\hline \noalign{\smallskip}
Surf.& Com-    & $R_0$  & $T$       & Glass   & $D$    \\
 No. & ments   & (mm)   & (mm)      &         & (mm)   \\
\hline 
 1  & Obsc  & $\infty$   & 189.472    & --        & 238.0 \\
 2  & L$_1$ & 2451.467   & 55.0       & FS        & 730.3 \\
 3  &       & -11435.65  & 219.271    & --        & 726.4 \\
 4  & Stop  & $\infty$   & 157.498    & --        & 660.2 \\
 5  & L$_2$ & -1425.622  & 55.0       & FS        & 656.7 \\
 6  &       & -1181.598  & 356.662    & --        & 662.5 \\
 7  & L$_3$ & -991.351   & 55.0       & FS        & 634.6 \\
 8  & Pri   & -2113.002  & -55.0      & Mirror    & 647.4 \\
 9  & L$_3$ & -991.351   & -356.662   & --        & 610.7 \\
 10 & L$_2$ & -1181.598  & -55.0      & FS        & 480.6 \\
 11 &       & -1425.622  & -157.498   & --        & 461.1 \\
 12 & SP    & $\infty$   & -219.271   & --        & 200.7 \\
 13 & L$_1$ & -11435.65  & -55.0      & FS        & 306.7 \\
 14 &       & 2451.467   & -119.919   & --        & 292.0 \\
 15 & L$_4$ & -330.389   & -35.210    & N-BK7     & 220.0 \\
 16 &       & 619.249    & -2.274     & --        & 216.1 \\
 17 & L$_5$ & 592.307    & -32.069    & N-BAK4    & 213.5 \\
 18 &       & -4564.917  & -20.998    & --        & 190.9 \\
 19 & F     & $\infty$   & -5.0       & N-BK7     & 173.9 \\
 20 &       & $\infty$   & -25.0      & --        & 171.2 \\
 21 & W     & $\infty$   & -5.0       & N-BK7     & 149.9 \\
 22 &       & $\infty$   & -15.0      & --        & 147.3 \\
 23 & Ima   & $\infty$   &            &           & 134.5 \\
\noalign{\smallskip} \hline
\end{tabular}
 
\vspace*{5pt} 
{\small Notes to Table~10: \\ 
Obscuration on surface No.~1 of diameter 238.0~mm. \\
All surfaces are spheres.}
\end{table}

\begin{table}[h]    
\caption{VT-119j design with an aperture of 500~mm and $45^\circ$ 
field of view. The effective focal length is 1358~mm.
The design waveband is $0.45-0.85 \mu$m.}
\begin{tabular}{cccccc}
\hline \noalign{\smallskip} 
Surf.& Com-   & $R_0$       & $T$     & Glass   & $D$    \\
 No. & ments  & (mm)        & (mm)    &         & (mm)   \\
\hline 
 1   & L$_1$  & 2279.917  & 95.0      & FS      & 1000.0 \\
 2   &        & 2456.517  & 522.273   & --      & 940.2  \\
 3   & L$_2$  & -4963.27  & 74.860    & FS      & 536.4  \\
 4   &        & -4531.56  & 0.001     & --      & 496.9  \\
 5   & Stop   & $\infty$  & 746.412   & --      & 491.0  \\
 6   & L$_3$  & -1044.83  & 101.742   & FS      & 1004.7 \\
 7   &        & -1075.10  & 75.840    & --      & 1081.8 \\
 8   & L$_4$  & -912.495  & 105.015   & FS      & 1109.1 \\
 9   &        & -1039.24  & 1637.765  & --      & 1209.4 \\
10   & Pri    & -2788.70  & -1431.74  & Mirror  & 2500.0 \\
11   & Ima    & -1328.68  & --        & --      & 1039.0 \\
\noalign{\smallskip}\hline
\end{tabular} 

\vspace*{5pt}
{\small Note to Table~11: \\ 
All surfaces are spheres.} 
\end{table}

\end{document}